\documentclass[10pt]{iopart}
\usepackage{graphicx}
\usepackage[utf8]{inputenc}
\newcommand\apj{Astrophys. J.}
\newcommand\aap{Astron. \& Astrophys.}
\newcommand\prl{Phys. Rev. Lett.}
\newcommand\apjl{Astrophys. J. Lett.}
\newcommand\mnras{Mon. Not. R. Astron. Soc.}
\newcommand\aapr{Astron. \& Astrophys. Rev.}
\newcommand\nat{Nature}

\newcommand\aj{Astron. J.}

\newcommand\ssr{Space Science Reviews}
\newcommand\araa{Annu. Rev. Astron. Astrophys.}
\newcommand\pasj{Publ. Astron. Soc. Jpn.}
\begin{document}

\title[Plasma heating and particle acceleration in astrophysical shocks]{Plasma heating and particle acceleration in collisionless shocks through astrophysical observations}
\author{M Miceli$^{1,2}$}
\address{$^1$ Universit\`a degli Studi di Palermo, Dipartimento di Fisica e Chimica E. Segr\`e, Piazza del Parlamento 1, 90134 Palermo, Italy}
\address{$^2$ INAF-Osservatorio Astronomico di Palermo, Piazza del Parlamento 1, 90134 Palermo, Italy}

\ead{marco.miceli@unipa.it}

\begin{abstract}
Supernova remnants (SNRs), the products of stellar explosions, are powerful astrophysical laboratories, which allow us to study the physics of collisionless shocks, thanks to their bright electromagnetic emission.
Blast wave shocks generated by supernovae (SNe) provide us with an observational window to study extreme conditions, characterized by high Mach (and Alfv\'enic Mach) numbers, together with powerful nonthermal processes.
In collisionless shocks, temperature equilibration between different species may not be reached at the shock front. In this framework, different particle species might be heated at different temperatures (depending on their mass) in the post-shock medium of SNRs.
SNRs are also characterized by a broadband nonthermal emission stemming at the shock front as a result of nonthermal populations of leptons and hadrons. These particles, known as cosmic rays, are accelerated up to ultrarelativistic energies via diffusive shock acceleration. If SNRs lose a significant fraction of their ram energy to accelerate cosmic rays, the shock dynamics should be altered with respect to the adiabatic case. This shock modification should result in an increase of the total shock compression ratio with respect to the Rankine-Hugoniot value of 4.
Here I show that the combination of X-ray high resolution spectroscopy (to measure ion temperatures) and moderate resolution spectroscopy (for a detailed diagnostic of the post-shock density) can be exploited to study both the heating mechanism and the particle acceleration in collisionless shocks. I report on new results in the temperatures measured for different ion species in the remnant of the SN observed in 1987 in the Large Magellanic Cloud (SN 1987A). 
I also discuss evidence of shock modification recently obtained in the remnant of SN 1006 a. D., where the shock compression ratio increases significantly as the angle between the shock velocity and the ambient magnetic field is reduced. 

\end{abstract}

\vspace{2pc}

\submitto{\PPCF}
\ioptwocol

\section{Introduction}
\label{intro}

Astrophysical shocks are ubiquitous in the universe, showing a wide variety of Mach numbers and characteristic length-scales. Astrophysical shocks are typically collisionless, because they expand in a rarefied medium, where Coulomb collisions cannot provide the viscous dissipation at the shock front. Collisionless shocks have been indeed observed at all scales, ranging from small, ``local'' shocks in the solar wind (\cite{ts85}) up to giant, cosmological shock waves (\cite{bdd08}).

Supernova remnants (SNRs) are the leftovers of supernova (SN) explosions, and are  characterized by powerful collisionless shocks. Their birth is associated with a sudden (less than one second) release of a huge amount of energy, which is of the order of  $E=10^{53}$ erg for core-collapse SNe and $E=10^{51}$ erg for Type Ia SNe (associated with the explosion of a white dwarf). Both core-collapse and Type Ia supernova explosions determine the violent ejection of supersonic stellar fragments (ejecta), having a characteristic kinetic energy of $10^{51}$ erg (which correspond to the energy that the Sun would radiate if it would maintain its current luminosity for almost 16 billion years, i.e. more than the age of the universe). The velocity of the ejecta can reach values as high as a few $10^4$ km s$^{-1}$ and their expansion and deceleration in the interstellar medium (ISM) drive strong shocks, which compress and heat the ambient medium (forward shocks) and the ejecta themselves (reverse shocks) up to X-ray emitting temperatures. 

\begin{figure*}[!ht]
\centering
\includegraphics[angle=0,width=\textwidth]{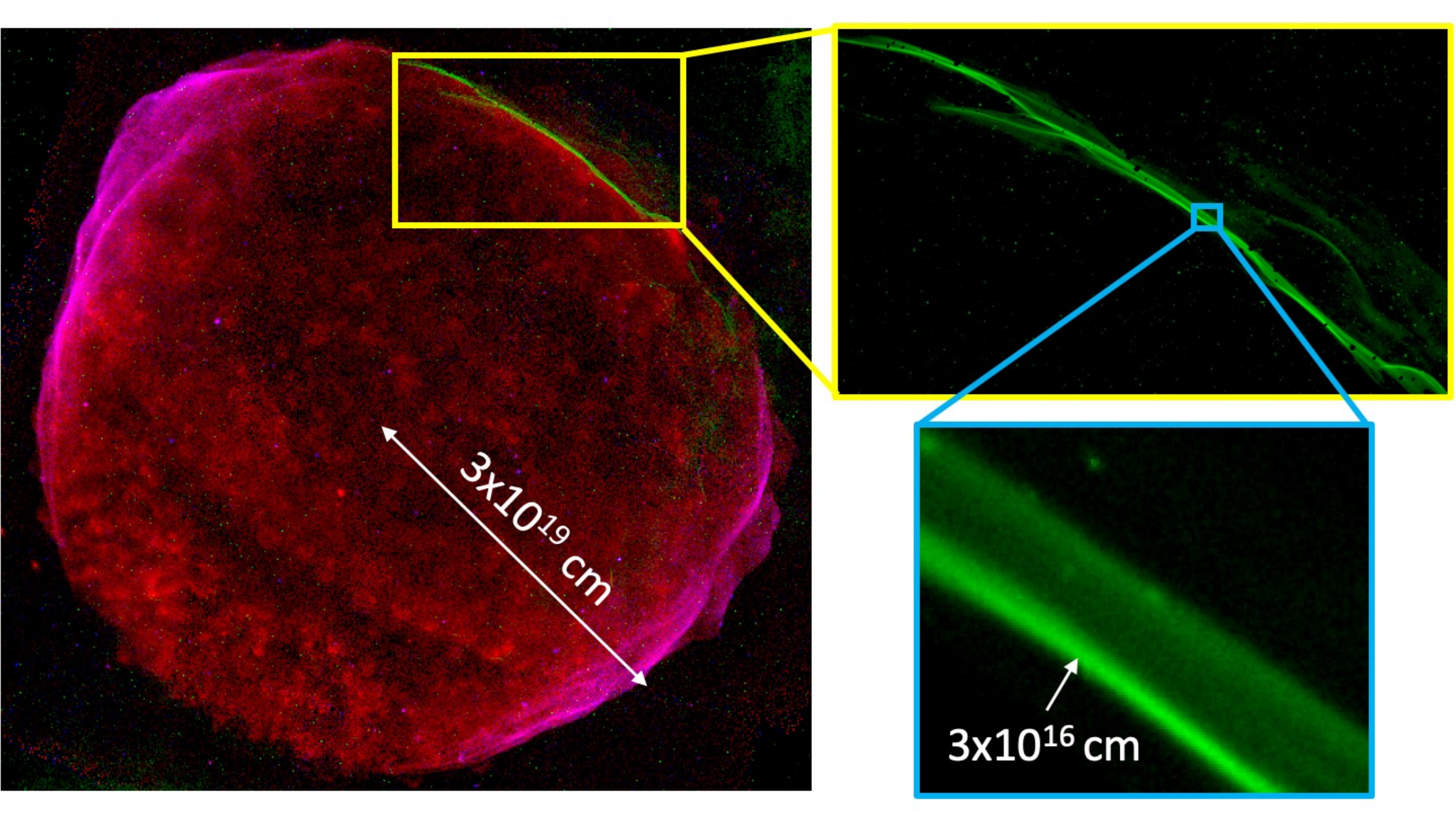}
\caption{\emph{Left panel}: Flux images of SN 1006 in the $0.5-1$ keV (red) and $2.5-7$ keV (blue) bands, obtained with the \emph{Chandra X-ray telescope}, together with Balmer H$\alpha$ emission (green). \emph{Right panels:} close-up views of the H$\alpha$ filament marking the shock front. In all panels North is up and East is to the left. Physical scales (indicated by white arrows) were obtained for a distance of $2.2$ kpc (\cite{wgl03}).}
\label{fig:sn1006}
\end{figure*}

X-ray observations of SNRs show spectacular extended nebulae, which reveal the complex morphology of shocked ISM and ejecta, and convey a wealth of information on many aspects of the remnant origin and evolution, and on the shock physics. Figure \ref{fig:sn1006} shows a composite image of the remnant of the supernova exploded in 1006 A. D. (hereafter SN 1006). Soft X-rays ($0.5-1$ keV, in red) show a complex pattern, with ripples of emission revealing knots of shocked ejecta (\cite{mbi09,wbg13}), hard X-rays ($2.5-7$ keV, in blue) are associated with synchrotron radiation in the northeastern and southwestern limbs (\cite{kpg95}), while optical H$_\alpha$ emission (in green) mark the northwestern shock front in great detail. By considering characteristic values for the post-shock temperature and density ($T=10^7$ K \cite{mbi09}, and $n=0.1$ cm$^{-3}$ \cite{mbd12,gmc22}, respectively), the Coulomb collision mean free path is of the order of $10^{19}$ cm, i. e. comparable with the radius of the remnant and much larger than the width of shock front. Since the viscous dissipation at the shock front is localized in an extremely narrow region, it cannot be provided by collisions and collective effects, as electromagnetic fluctuations and plasma waves, instead, are thought to be responsible of the plasma heating. Indeed, in this case the shock width might be of the order of $10^8$ cm, given that the ion inertial length for SN 1006 is of approximately $10^8$ cm, and the proton gyroradius for protons with the same speed as the shock ($5000$ km/s \cite{wbg13}) in a magnetic field of $90$ $\mu$G (\cite{mab10,bal06}) is of approximately $5\times10^8$ cm. 
 
The sudden heating and compression within the thin shock front is followed by a slow relaxation toward the equilibrium in the downstream flow. Therein, the almost neutral optically thin coronal plasma is slowly ionized through electron-ion collisions, and reaches the collisional ionization equilibrium only after a timescale of approximately $t_{CIE}=10^{12}/n$ s (where $n$ is the post shock density in cm$^{-3}$, \cite{sh10}), which corresponds to more than $10^5$ yr for the typical densities of SN 1006, i. e., more than ten times the age of the remnant. It is then unsurprising that the X-ray spectra of SNRs typically show signatures of an underionized plasma, though with some notable exception (e. g., W49B \cite{oky09,mbd10,ytw18}, IC 443 \cite{kon05,gmo18,otu21} and W44 \cite{otu20}). 

Moreover, different particle species are expected to be heated at different temperature in collisionless shocks, and their thermalization proceeds slowly in the post shock medium. For example, the time-scale for electron-proton temperature equilibration is $t_{ep}\sim 10^4n^{-1}(kT/\rm{1~keV})^{3/2} (\ln{\Lambda}/31)$ yr, which can be as well larger (much larger for historical SNRs) than the age of a remnant. For example, in SN 1006, $t_{ep}\sim10^5$ yr. In general, therefore, the shocked plasma in SNRs is \emph{not} in equilibrium.

Shock fronts of SNRs are known to be powerful particle accelerators, with particles gaining energy by diffusively crossing back and forth the shock front through the first-order Fermi mechanism (or diffusive shock acceleration, DSA). Radio synchrotron radiation from ultrarelativistic (GeV) electrons is systematically observed in the shell of SNRs \cite{gre19}, and X-ray synchrotron emission from electrons with energies reaching $\sim 10^{13}$ eV can be observed at the shock front of young SNRs (\cite{vin12}), including SN 1006 (see Fig. \ref{fig:sn1006}).  Evidence for high energy hadrons in SNRs has been also obtained in a few cases (\cite{tgc10,mc12,aaa13,sle14}).

Here, I first focus on the shock heating mechanism and on the evolution of the ion temperature in the shocked plasma in the remnant of the supernova observed on 1987 February 23 in the Large Magellanic Cloud (hereafter SN 1987A). I will then show the effects of the back-reaction of hadron acceleration in the shock front of SN 1006. 

The paper is organized as follows: Section \ref{87a} is dedicated to the collisionless shock heating in SN 1987A, Sect. \ref{1006} shows the shock modification induced by efficient particle acceleration in SN 1006, while my conclusions are summarized in Sect. \ref{concl}

\section{Collisionless shock heating in SN 1987A}
\label{87a}
In a collisionless shock propagating in a medium with different particle species, by adopting a multi-fluid approach, one may be tempted to write down the Rankine-Hugoniot conditions for each particle species. In this framework, the post-shock temperature $T_i$ for the i-th species might be obtained from the relation:
\begin{equation}
    kT_i=\frac{3}{16}m_i v_{sh}^2
\label{eq:ts}
\end{equation}
where $m_i$ is the particle mass for the i-th species and $v_{sh}$ is the shock velocity. A mass-proportional post-shock temperature is expected in the case of scattering isotropization of the incoming particles by plasma waves. However, this approach does not consider that  (partial) equilibration between different species can also occur within the shock front (e.g., by wave-particle interactions). 

It is not straightforward to test the validity of eq. \ref{eq:ts} by considering protons and electrons. This is because the processes responsible for electron heating in collisionless shocks are expected to be different from those acting for protons and ions  (\cite{bdd08,sh00,pcs15}).
Indeed, the electron to proton temperature ratio in SNRs is lower than 1, though being typically larger than that predicted by eq. \ref{eq:ts} \cite{rgh03,gsm13,ray18,vhm08}, and showing a dependence on the shock velocity which can be described by $T_e/T_p\propto v_{sh}^{-2}$ (\cite{glr07}). 

To test whether in collisionless shocks the post-shock temperature scales linearly with the particle mass, it is then crucial to measure the temperature of ions with different masses in the shocked plasma. Early UV observations of SN 1006 indicated an Oxygen to proton temperature ratio larger than 1, but lower than that predicted by eq. \ref{eq:ts} (\cite{krz04}). On the other hand, more recent measurements have shown a mass-proportional heating for  He, C and N ions in the same remnant. Since bright emission line complexes of heavy ions are typically observed in the X-ray band, the analysis of X-ray spectra is crucial to extend the study of collisionless particle heating to a wider range of masses. 

The temperature of ions can be deduced by the thermal broadening of their X-ray emission lines (\cite{vlg03,bvm13}).
This is a delicate measurement, since different effects contribute to the line broadening of an emission line stemming in the plasma of a SNR, namely i) the intrinsic resolution of the X-ray spectrometers, ii) the bulk Doppler motion (if approaching and receding plasmas lie along the same line of sight); iii) the angular extension of the X-ray emission (current high resolution spectrometers are typically slitless, so the spectrum is convolved with the profile of the remnant emission in the dispersion direction), and iv) thermal broadening. It is then crucial to disentangle the contribution of thermal broadening from other effects to get an accurate measurement of the ion temperatures.

To this end, SN 1987A is a privileged target: its X-ray emission has been monitored regularly, thus providing a unique dataset for the evolution of the X-ray spectra in a young SNR. Moreover, X-ray data are complemented by a wealth of observations in an extremely wide range of wavelengths (from radio to $\gamma-$rays), which provided us with an accurate diagnostics of the complex interaction of the shock front with the surrounding inhomogeneous medium (\cite{mcr93, mcf16}). 
The constraints derived from the multi-epochs and multi-wavelengths observations were adopted as setup parameters to develop a thorough 3-D magnetohydrodynamic (MHD) model, which is able to describe in great detail both the origin and the evolution of the remnant and of its thermal and nonthermal emission self-consistently (\cite{omp16,omp19,onf20,oon20}).

A novel approach combining data and MHD simulations was developed to analyze the high resolution X-ray spectra of SN 1987A collected with the \emph{Chandra} X-ray telescope (\cite{mob19}). The work was focused on the 2007 and 2011 data, i. e., the two deepest observations performed with the \emph{Chandra} High Energy Transmission Grating (HETG) available at that time. The accurate comparison of the actual spectra with those synthesized from the MHD model, made it possible to isolate the contribution of thermal broadening to the observed line widths of several emission lines, thus deriving the post shock temperature of different heavy elements, namely of Ne, Mg, Si and Fe. 
Results clearly show that, for these species, the ion to proton temperature ratio is always significantly higher than one and is consistent with increasing linearly with the ion mass, as predicted by eq. \ref{eq:ts}, for a wide range shock parameters (\cite{mob19}).
A similar methodology was later adopted to analyze the Chandra HETG observation taken in 2018 (\cite{rpz21}), confirming that, on average, the observed line widths are well described when thermal broadening associated with a mass proportional heating is taken into account.

\begin{figure}[!ht]
\centering
\includegraphics[angle=0,width=\columnwidth]{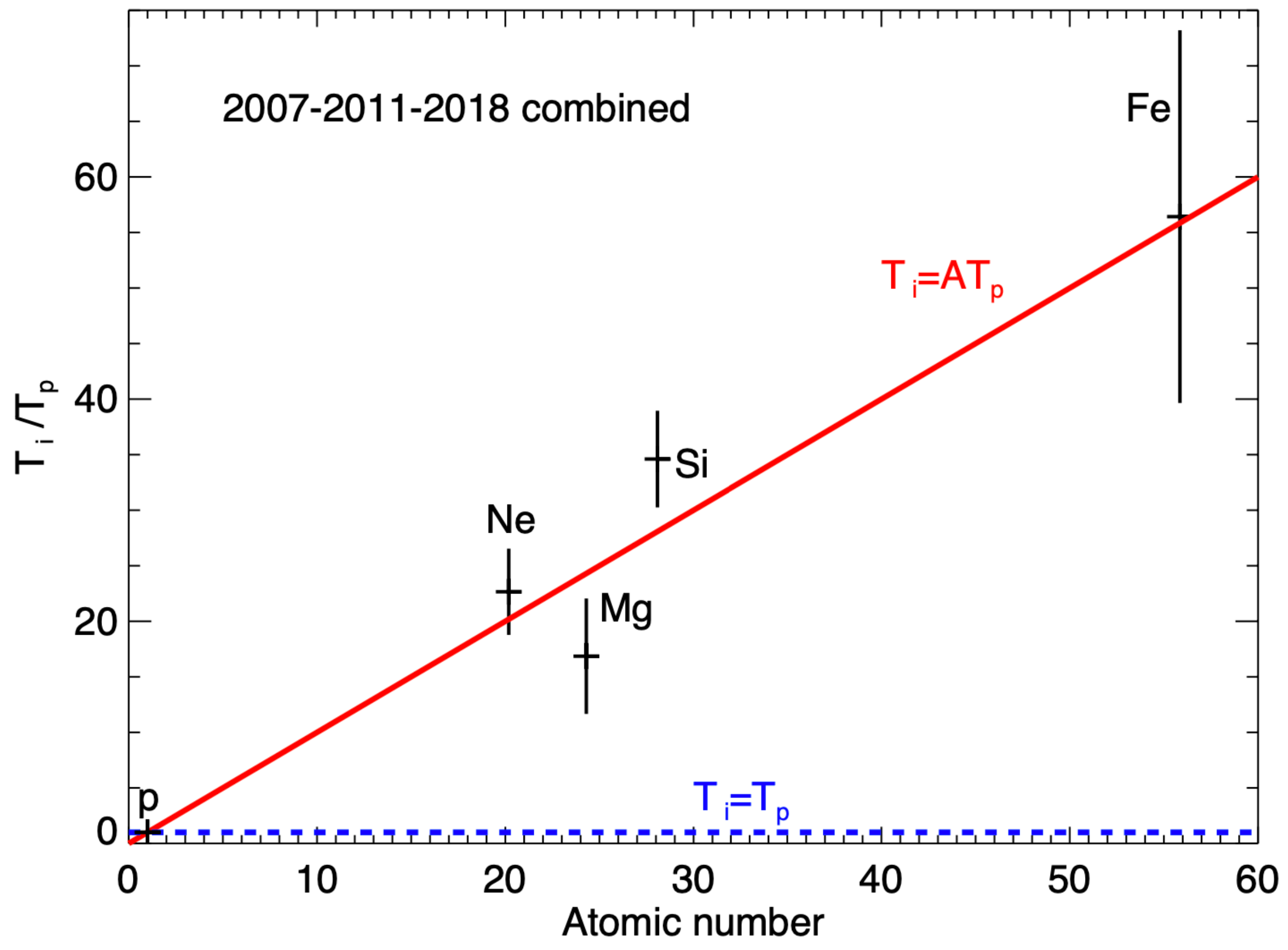}
\caption{Ion to proton temperature ratios for Ne, Mg, Si and Fe in SN 1987A obtained by combining the 2007 and 2011 results  (\cite{mob19}) with the 2018 results (\cite{rpz21}). The red line marks the mass-proportional trend predicted by eq. \ref{eq:ts}.}
\label{fig:titp}
\end{figure}

I here combine the results obtained for the 2007-2011 (\cite{mob19}) spectra with those obtained for the 2018 observation (\cite{rpz21}). By adopting the procedure described in \cite{mob19}, I compare the actual line widths with those predicted by the MHD simulation with and without including the effects of thermal broadening, thus determining the ion temperature. Figure \ref{fig:titp} shows the ion to proton temperature ratio for Ne, Mg, Si and Fe, together with the mass-proportional relation predicted by eq. \ref{eq:ts}. The general trend clearly confirms previous results and is in remarkable agreement with predictions of hybrid simulations of collisionless shocks, which shows that the post shock temperature scales linearly with the atomic mass $A$ (with simulations performed up to A=8, \cite{cys17}).

On the other hand, Fig. \ref{fig:titp} shows that the post-shock temperature of Mg is slightly lower than expected. The measure of the Mg temperature mainly depends on the width of its brightest emission line, namely the Ly$\alpha$ for the ionization state Mg XII at $\sim8.4$ \AA. A possible explanation for the relatively low temperature of Mg is that this emission line stems well behind the shock front, in a region where ions are approaching thermal equilibrium with other particles and the temperature is expected to be lower than that achieved at the shock front. Temperature equilibration between ions and protons is expected to be reached when the time integral of the electron density reaches $\tau \sim5-10\times10^{10}$ s cm$^{-3}$ (\cite{spi62}). On the other hand, it must be considered that the thermalization timescale increases by about a factor of 5 (\cite{nm01}) in a very turbulent magnetic field, which is expected to be present in the SN 1987A shocked medium \cite{zsg18}. This suggests that significant temperature variations with respect to the immediate post-shock region are expected for those ions whose emission lines originate in a plasma with $\tau$ larger than $\sim2-5\times10^{11}$ s cm$^{-3}$. The MHD simulations adopted to synthesize the spectra include Coulomb collisions between electrons and protons and follow their temporal evolution downstream, while collisions between ions and protons are not included. However, thanks to the synthetic spectra, it is possible to derive the value of $\tau$ in the plasma which mainly contribute to the line emission of a given element. I here focus on the synthetic spectra extracted from model B18.3 (\cite{oon20}, see also \cite{gmo22} for a similar approach for the Fe K line emission) describing SN 1987A 31 years after the explosion (corresponding to 2018 a. D.). 

\begin{figure*}[!ht]
\centering
\includegraphics[angle=0,width=\textwidth]{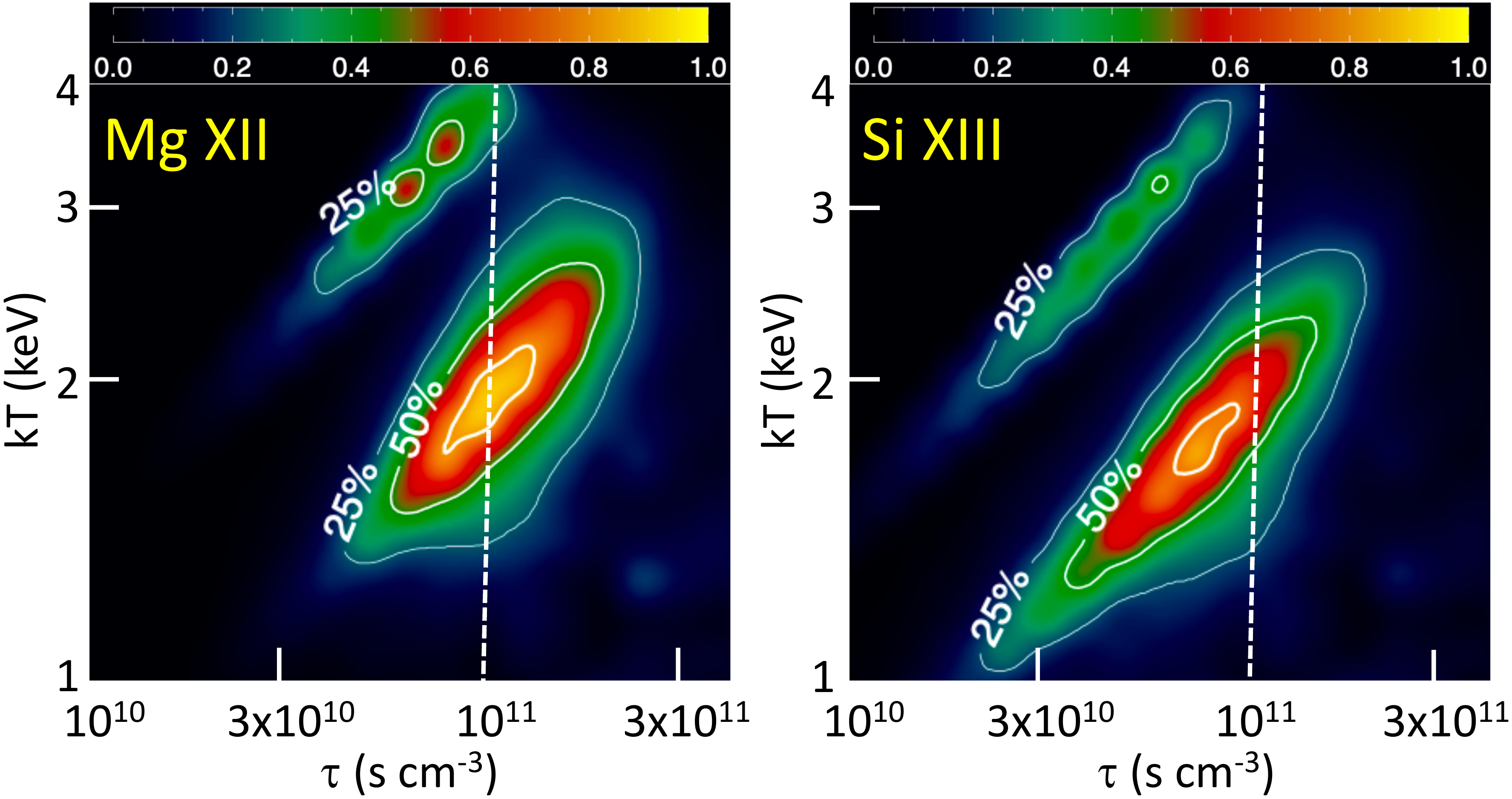}
\caption{Normalized continuum-subtracted line flux of the Mg XII Ly$\alpha$ (\emph{left panel}) emission line and of the Si XIII He$\alpha$ (\emph{right panel}) emission line as a function of the plasma temperature and time integral of the electron density ($\tau$). Contours mark 25\%, 50\% and 75\% of the maximum.}
\label{fig:em}
\end{figure*}

Figure \ref{fig:em} shows how the different values of temperature and ionization parameter of the X-ray-emitting plasma in SN 1987A contribute to the line emission. In particular, left panel of Fig. \ref{fig:em} shows the normalized continuum-subtracted line flux of the Mg XII emission line as a function of the plasma temperature and $\tau$. The figure clearly shows that the bulk of the emission line originates in a region where $\tau$ is slightly larger than $10^{11}$ s cm$^{-3}$, i. e. where thermalization between Mg and colder protons starts to be relevant. This is in nice agreement with the Mg showing a lower temperature than that expected at the shock front. As a comparison, I also report the distribution of the continuum-subtracted line flux for the Si XIII emission line (He$\alpha$ at $6.65$ \AA, right panel of Fig. \ref{fig:em}), which is the brightest Si line (i.e., the estimate of the Si temperature mainly depends on its width): in this case the bulk of the line emission originates from a plasma with $\tau<10^{11}$ s cm$^{-3}$, where we expect the ion temperature to be almost the same as that achieved at the shock front. 

Further investigations require the inclusion of ion-proton collisions in the model. However, this approach already shows that the combination of data analysis and MHD simulations provides a powerful diagnostic tool to study both the immediate post-shock conditions and the evolution toward the equilibrium in the post shock flow.

\section{Shock modification in SN 1006}
\label{1006}

As explained in Sect. \ref{intro}, collisionless shock heating is associated with particle acceleration and SNRs are known to be sites of efficient acceleration.
Self consistent hybrid (kinetic ions-fluid electrons) simulations show that the acceleration efficiency is expected to increase by reducing the angle between the  shock velocity and the ambient magnetic field and that magnetic turbulences are associated with efficient ion
acceleration (quasi-parallel scenario, e. g., \cite{cs14}). This is in agreement with measurement of radio polarization. For example, the radio polarization study of SN 1006 shows efficient particle acceleration and higher magnetic turbulence for quasi-parallel shocks (\cite{rhm13}). Similarly, radio polarization measurements in Kepler's SNR indicate an almost radial (i. e. parallel to the shock velocity) magnetic field and a lower level of polarization (\cite{dkr02}), which is indicative of high magnetic turbulence in the acceleration site (\cite{smb22}). On the other hand, re-acceleration of ambient cosmic-ray seeds is nearly independent on the shock inclination (\cite{czs18})

Indeed, SNRs can sustain the observed flux of Galactic cosmic rays provided that they inject $10-20\%$ of their kinetic energy to the particles (e. g., \cite{hil05}). Such effect, if present, is expected to alter the shock dynamics by increasing the shock compression ratio, and decreasing of the post-shock temperature with respect to the Rankine-Hugoniot values (\cite{dru83,deb00,bla02,vyh10,sle14}). In particular, recent hybrid simulations show the formation of a shock postcursor associated with efficient particle acceleration. This postcursor moves downstream (approximately at the Afv\'en speed), acting as an additional energy sink, which allows non-linear magnetic fluctuations, and particles, to drift away from the shock front (\cite{hc20}). When the effects of the postcursor are taken into account, a shock compression ratio $r_s\sim6-7$ (to be compared with the canonical value of 4 for strong shocks) can be achieved when the cosmic ray pressure is of the order of 10\% of the bulk ram pressure.

The presence of shock modification has been recently revealed in SN 1006 \cite{gmc22}. In this remnant, thanks to the the almost uniform magnetic field, approximately aligned in the southwest-northeast direction, it is possible to observe, in the same object, regions with efficient particle acceleration (i.e. regions in quasi-parallel conditions, indicated by the blue limbs in Fig. \ref{fig:sn1006}) and regions where we do not expect shock modification (\cite{mbd12}). A careful spatially resolved spectral analysis of the X-ray spectra stemming from the shocked interstellar medium at different positions of the shell reveals a regular azimuthal modulation of $r_s$, with a minimum $r_s=4$ in quasi-perpendicular conditions and a maximum which reaches values as high as $r_s=7$ in quasi-parallel conditions (\cite{gmc22}). The accurate comparison of the azimuthal profile of the shock compression ratio with state of the art theoretical models of modified shocks including the effect of the postcursor shows a nice agreement between model and observations and provide important constraints on the acceleration efficiency and on its dependence on the angle between the ambient magnetic field and the shock velocity. In particular, in quasi-parallel conditions, the cosmic rays pressure is $\sim12\%$ of the total, while the normalized magnetic pressure is $\sim5\%$. Moreover, the $r_s$ azimuthal profile shows a sharp minimum, which is strongly indicative of efficient (normalized pressure $\sim6\%$) re-acceleration of pre-existing Galactic cosmic rays (\cite{gmc22}). 

The shock modification is expected to affect the spectrum of the accelerated particles, by modifying their spectral index (\cite{chb20}). Remarkably, when considering the model parameters that best reproduce the azimuthal profile of $r_s$, a spectral index of $\sim 2.2$ is obtained for the energy spectrum in the regions with maximum acceleration efficiency (the nonthermal limbs). When considering the synchrotron emission from the accelerated electrons, this corresponds to a photon index $\alpha\sim0.6$, which is in nice agreement with that observed for SN 1006 (\cite{gre19}).

\section{Conclusion}
\label{concl}

Astrophysical shocks are natural laboratories which allow us to access the shock physics at extreme conditions.
X-ray observations of SNRs are a powerful diagnostic tool to study the abrupt collisionless shock heating and the relaxation of the post shock plasma towards the equilibrium. Also, X-ray observations can allow us to observe synchrotron radiation from ultrarelativistic electrons accelerated at the shock front and to probe the back-reaction of cosmic rays on the shock dynamics (\cite{syb22}). 

To extract all the information stored in the X-ray data, a detailed comparison with state of the art models is necessary. 

The physical processes localized at the shock front affect the whole remnant structure and its global evolution. Therefore, both hybrid simulations investigating the microphysics at the shock front and MHD simulations describing the evolution of the system on larger spatial and time scales provide  powerful tools to get a higher level of diagnostics.

Moreover, the upcoming generation of X-ray telescopes (as \emph{XRISM} and \emph{Athena}) will be equipped with high resolution spectrometers, based on microcalorimeters, which  will provide us with an unprecedented level of details. The development of novel advanced tools for data analysis and for quantitative comparison with theoretical models is a challenging task which will allow us to exploit the quality of the data in the near future.

\section*{Acknowledgements}
This work was partially supported by the INAF mini-grant ``X-raying shock modification in supernova remnants".

\section*{References}
\providecommand{\newblock}{}

\end{document}